\documentclass[prd,eqsecnum,showpacs,nofootinbib]{revtex4}
\usepackage{bm}
\usepackage{epstopdf}
\usepackage{amsmath,amsfonts,amssymb}
\usepackage{graphicx}
\usepackage[colorlinks=true, allcolors=blue]{hyperref}

\begin{document} 
 
\title{Properties of Quantum Reactivity\\ for a Multipartite State}

\author{Shahabeddin~Mostafanazhad~Aslmarand$^{1}$, Warner A. Miller$^{1}$,\\ Tahereh Razaei$^{1}$,
Paul~M.~Alsing$^{2}$ \& Verinder~S.~Rana$^{3}$}

\affiliation{
$^{1}$Department of Physics, Florida Atlantic University, Boca Raton, FL, 33431, USA\\
$^{2}$Air Force Research Laboratory, Information Directorate,Rome, NY 13441, USA\\
$^{3}$Naval Information Warfare Center Pacific (NIWC PAC), San~Diego, CA 92152, USA
}

\begin{abstract}
We discuss the properties of quantum state reactivity as a measure for quantum correlation. This information geometry--based definition is a generalization of the two qu$b$it construction of Schumacher to multipartite quantum states.  It requires a generalization of information distance to information areas as well as to higher--dimensional volumes. The reactivity is defined in the usual chemistry way as a ratio of surface area to volume. The reactivity is an average over all detector settings. We show that this measure posses the key features required for a measure of quantum correlation.  We show that it is invariant under local unitary transformations, non--increasing under local operations and classical communication, and  monotonic.  Its maximum bound can't be obtained using only classical correlation.  Furthermore, reactivity is an analytic function of measurement probabilities and easily extendable to higher multipartite states.  
\end{abstract} 

\pacs{03.67.a,03.65.Ud,03.65.w}

\maketitle

\section{Information Geometry in Quantum Mechanics}
\label{sec:intro}
The central role that information plays is in physical laws are well established and captured by two pithy phrases, ``Information is Physical," and ``It--from--Bit.''\cite{Wheeler:1990,Landauer:1991, Zurek:1989}  Within this framework, Schumacher introduced a triangle inequality that is based on measurements of a singlet state.\cite{Schumacher:1991}  His approach was an innovative application of  quantum information geometry that highlighted quantum entanglement between two qu$b$its.  Schumacher's original construction for the singlet state was based on the quantum information distance measure of Rokhlin and Rajski.  This distance formula was first used in quantum mechanics by Zurek and Bennett et al.\cite{Rokhlin:1967,Rajski:1961,Zurek:1989,Bennett:1998}

Quantum entanglement is the key resource for quantum information processing.\cite{Preskill:2012}  This has been accompanied by a wealth of studies of entanglement and entanglement measures.  Here we report on the properties of a recent information geometry measure of entanglement  arising from an extension of  Schumacher's construction from bipartite to multipartite states\cite{Miller:2018,Aslmarand:2019}   In particular, we introduced a geometric-based measure of {\em reactivity} that is a ratio of surface area to volume.\cite{QIG:1990,Miller:2019}

We show that the quantum state reactivity satisfies the properties required for such a measure of quantum correlation. We discuss these properties in Sec.~\ref{sec:properties}.  However, we will first provide a brief outline of Schumacher's information distance--based geometry in Sec.~\ref{sec:Schumacher} in order to motivate our  generalization to multipartite quantum states.  In Sec.~\ref{sec:Schumacher}, we also extend this distance to information area and volumes.  These higher--dimensional volumes can be used to define the {\em reactivity} measure for quantum correlation. The reactivity is directly related to the definition used in chemistry as the ratio of surface area divided by volume.  This is easily generalizable to higher dimensional multipartite quantum states. In Sec.~\ref{fini} we discuss and summarize our results. 

\section{Definition of Mean Reactivity to Generalized Multipartite Quantum States}
\label{sec:Schumacher}

Schumacher showed that geometries created for entangled quantum states are not ordinarily embeddable in Euclidian space.\cite{Schumacher:1991}   He utilized an information metric based on Shannon entropy as:
\begin{equation}
 \label{Eq:D}
 {\mathcal D}_{AB} = H_{A|B} + H_{B|A} = 2H_{AB}-H_{A}-H_{B},
 \end{equation}
where $H_{A|B} $ is conditional entropy of $A$ given $B$.\cite{Rokhlin:1967,Rajski:1961,Zurek:1989,Bennett:1998}

\begin{figure}[h!]
\centering
\includegraphics[width=3 in]{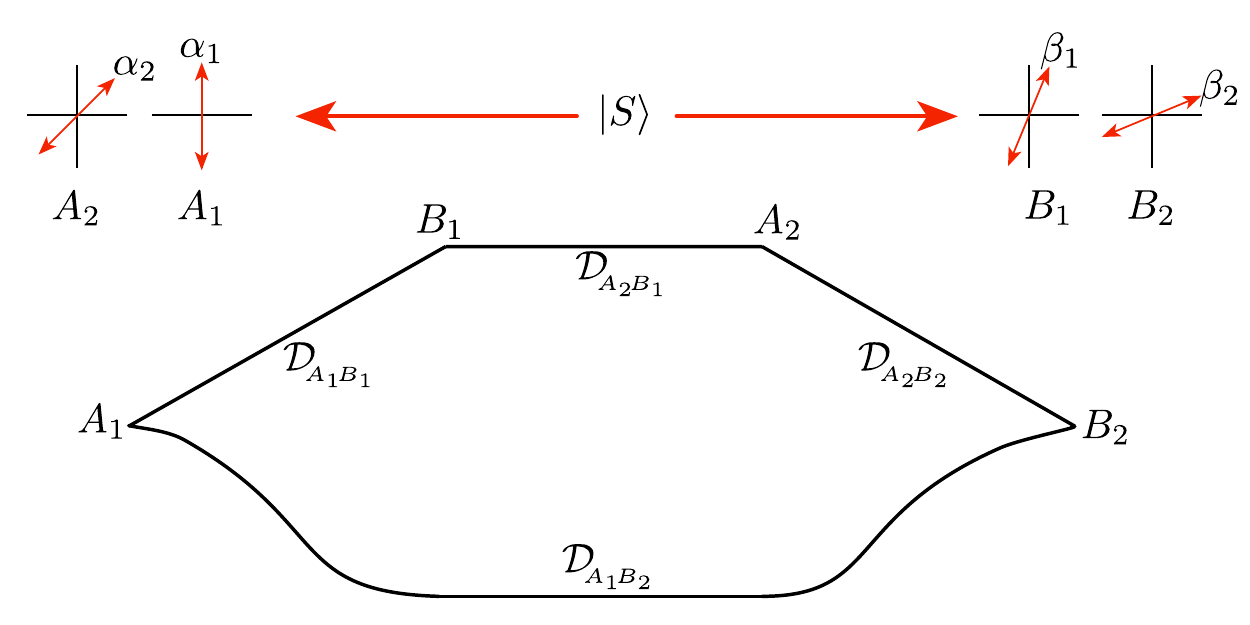}
\label{fig:Schumacher}
\caption{The trapezoidal information geometry of an ensemble of singlet states, $|S\rangle$, that was first introduced by Schumacher.\cite{Schumacher:1991} Here Alice employs two detectors $A_1$ and $A_2$ measuring one of the two entangled photons, and Bob also has two distinct detectors $B_1$ and $B_2$. Alice and Bob record a ``1'' if their detectors triggers, otherwise they measure a ``0.''  From their string of measurements they can define an  information distance ${\mathcal D}_{A_iB_j}$ between each of the four pairs of detectors.  Schumacher found that for some choice of detector angles, namely
$\{\alpha_1,\beta_1,\alpha_2,\beta_2\} =\{0,\pi/8,\pi/4,3\pi/8 \}$,
that the trapezoidal geometry could not be embedded into the Euclidean plane. The triangle equality would be violated for this entangled state. }
\end{figure}

This measure of distance satisfies all three properties of a metric,
\begin{enumerate}
\item It is constructed so as to be symmetric, ${\mathcal D}_{AB} = {\mathcal D}_{BA}$
\item It obeys the triangle inequality, ${\mathcal D}_{AB}\geq {\mathcal D}_{AC}+ {\mathcal D}_{CB}$.
\item It is non-negative, ${\mathcal D}_{AB}\geq 0$, and equal to 0 when $A$“=”$B$.
\end{enumerate}
In order to connect this distance to a measure of quantum correlation, Schumacher created a trapezoidal structure for a bipartite system, by assigning two detector for each of the two observers measuring each qubit as illustrated in Fig.~\ref{fig:Schumacher}. 

In this geometry, and for a range of detector settings, the direct distance between the two detectors $A$ and $D$ is larger than the sum of the three indirect distances.  In other words, Schumacher showed that the inequality can be violated for maximally entangled states.
\begin{equation}
\label{eq:D}
{\mathcal D}_{A_1B_2} \leq{\mathcal D}_{A_1B_1}+{\mathcal D}_{A_2B_1}+{\mathcal D}_{A_2B_2}.
\end{equation}
This is equivalent to the non-embeddibility of the trapezoid into the Euclidean plane. In this simple
gedanken experiment,  he showed that it's possible to capture the none-classicality of the quantum correlation of a quantum system by looking at the Shannon-based information geometry applied directly to the space of measurements obtained by the four detectors of Alice and Bob. 

\begin{figure}[h!]
\centering
\includegraphics[width=5 in]{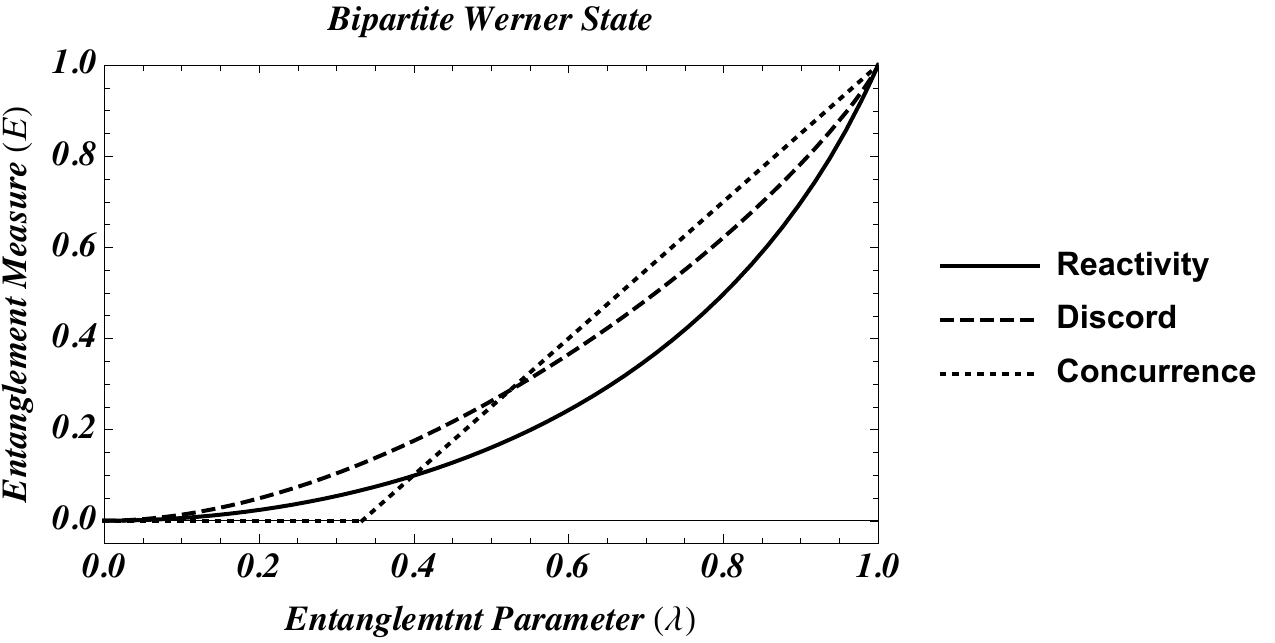}
\label{fig:cdr}
\caption{Comparison of concurrence and discord with our definition of reactivity for a bipartite Werner state. Concurrence provides a measure for entanglement in that it is zero for separable states; however, it may be difficult to implement  for higher--dimensional multipartite states.  Both global quantum discord and reactivity are measures for quantum correlation and not entanglement. Global quantum discord will always be an upper bound for reactivity; however, it may increase under LOCC in some cases.  Reactivity is non-increasing under LOCC.\cite{Vedral:2017}   }
\end{figure}

Following Schumacher's approach, we generalized this approach to a multipartite system containing an arbitrarily large number of qu$b$its.\cite{Miller:2018,Aslmarand:2019,Miller:2019}  In so doing, we defined a reactivity $\mathcal{R}$ for a  quantum network, and for a bipartite quantum state it is expressed in terms of the information distance, 
\begin{equation}
\label{eq:R1}
{\mathcal R}:=\frac{1}{\overline{{\mathcal D}_{AB}}}.
\end{equation}
Here, the average is taken over all detector settings,  
\begin{equation}
\label{eq:Dbar}
\overline{D_{AB}} := \frac{1}{4\pi^2} \int_0^{2\pi}  \int_0^{2\pi}  D_{AB}(\alpha,\beta) d\alpha d\beta.
\end{equation}
In Fig.~\ref{fig:cdr} we compare the reactivity ($\mathcal R$) to the other commonly--used measures, concurrence ($\mathcal C$)\cite{Wootters:1998,Rungta:2003} and discord ($\mathcal D$)\cite{Zurek:2001,Rulli:2011}  for a bipartite Werner state,
\begin{equation}
|W\rangle=\lambda |S\rangle\langle S | + \frac{1}{4} \left(1-\lambda\right)I,
\end{equation}  
where $\lambda\in [0,1]$ is the entanglement, and $|S\rangle=(|00\rangle+|11\rangle)/\sqrt{2}$ is a singlet state. 

In addition to information distance in Eq.~\ref{eq:D}, we can analogously assign an information area of a tripartite quantum state\cite{QIG:1990}, where  
\begin{equation}
\label{Eq:A}
{\mathcal A}_{ABC} := H_{A|BC}H_{B|CA}+H_{B|CA}H_{C|AB}+H_{C|AB}H_{A|BC}.
\end{equation}
This can be extended analogously to higher-dimensional simplexes, e.g.  the information volume for a 4--qubit quantum state\cite{QIG:1990,Miller:2018,Aslmarand:2019} such that 
\begin{equation}
\label{Eq:V}
\begin{array}{ll}
{\mathcal V}_{ABCD}& := H_{A|BCD}H_{B|CDA}H_{C|DAB} +H_{B|CDA}H_{C|DAB}H_{D|ABC}\\
&\ \ +H_{C|DAB}H_{D|ABC}H_{A|CDB}+H_{D|ABC}H_{A|BCD}H_{B|CDA}.
\end{array}
\end{equation}
Such higher--dimensional volumes enable us to define the reactivity,  
\begin {equation}
\label{Eq:C}
{\mathcal R} := 
\left(
\frac{   \overline{  {}^{(d\!-\!2)}\!Area  }  }{  \overline{  {}^{(d\!-\!1)}\!V\!olume. }   }
\right)
,
\end{equation}
for higher number of qu$d$its.  Here for qu$d$it state the volume is $(d\!-\!1)$-dimensional volume and the area is its $(d\!-\!2)$--dimensional boundary.  
We showed that for Werner state that reactivity increases as quantum correlation increases.\cite{Aslmarand:2019}  In this paper, we suggest that this reactivity satisfies the requisite properties for a measure of quantum correlation.

\section{Mean Reactivity as Candidate Measure of Quantum Correlation}
\label{sec:properties}
We show in this section that this measure of mean reactivity ${\mathcal R}$ satisfies the four established properties for a measure of quantum correlation.\cite{Bennett:1996}. Our proposed geometrical measure of correlation satisfies the following properties:
\begin{enumerate}
\item Reactivity is invariant under unitary transformations.
\item Reactivity  is non-increasing under LOCC's.
\item Reactivity is a monotonic function in quantum correlation, and the maximum bound on this curvature can't be obtained using only classical correlation.
\end{enumerate}
We address each of these in order in the next four subsections, Sec.~\ref{sec:property1}-Sec.~\ref{sec:property3}.  Its very definition based on the ratio of analytic functions shows that it is extendable to higher--dimensional multipartite quantum states. 

\subsection {Invariance under Local Unitary Operators} 
\label{sec:property1}
We know that the reactivity defined in Eq.~\ref{eq:R1} depends on the information distance which has the following properties
\begin{equation}
\begin{split}
D_{AB} &= H_{A|B} + H_{B|A}\\
&= Tr\left(M_A \otimes M_B \rho\right)\, \log\left[ \frac {Tr(M_A \otimes M_B\, \rho) ^2}{Tr\left(M_A Tr_B (\rho)\right) \, Tr\left(M_B Tr_A(\rho)\right)}\right]
\end{split}
\end{equation}
Which $M_A$ and $M_B$ are the projecton operators for Alice and Bob, now if we apply a unitary transformation to initial state 
\begin{equation}
\rho \longrightarrow U_A \otimes U_B \, \rho \, {U_A}^T \otimes {U_B}^T 
\end{equation}
then 
\begin{equation}
\begin{split}
D_{AB} &= Tr\left(M_A \otimes M_B \left(U_A \otimes U_B \, \rho \, {U_A}^T \otimes {U_B}^T \right)\right) \\
&\log{\left[ \frac {Tr\left(M_A \otimes M_B \left(U_A \otimes U_B \, \rho \, {U_A}^T \otimes {U_B}^T \right)\right) ^2}{Tr\left(M_A U_A Tr_B (\rho){U_A}^T\right) \,Tr\left(M_B {U_B} Tr_A(\rho){U_B}^T \right)}\right]}\\
                  & =   Tr\left( \left({U_A}^T \otimes {U_B}^T M_{A} \otimes M_B U_A \otimes U_B \right) \, \rho  \right)\\
                  &\log\left[ \frac { Tr\left( \left({U_A}^T \otimes {U_B}^T M_{A} \otimes M_B U_A \otimes U_B \right) \, \rho  \right)^2}{Tr\left({U_A}^TM_A U_A Tr_B (\rho)\right) \, Tr\left({U_B}^T  M_B {U_B} Tr_A(\rho)\right)}\right]\\
\end{split}
\end{equation}
Any arbitrary $2\times 2$ unitary quantum gate can be written as  a phase shift multiplied by a rotation.  This rotation can be expressed as a composition of a rotation and a rotation about the z-axis, 
\begin{equation} 
\begin{split}
&M_A = \sum  \lambda_a  |a \rangle \langle a |\\
&{U_A}^TM_A U_A = \sum \lambda_a  R(z)^T R(\alpha)^T |a \rangle \langle a | R(z) R(\alpha)
\end{split}
\end{equation}
In other words, we rotate the detector of Alice by some fixed angle about some axis for all measurements.  Likewise, we rotate the detector of  Bob independently by some other fixed angle and other rotation axis. This operation will not change $\kappa$ since we took an average of the information distance over all possible configurations of Alice and Bob in EQ.~\ref{eq:Dbar}.
Then unitary LOCC's will not change our measure of correlation.

\subsection{Reactivity is Non-Increasing under LOCC's} 
\label{sec:property2}
It is accepted that any definition of a measure for quantum correlations that it should not increase under any LOCC.  We argue that for our reactivity measure in  Eq.~\ref{eq:R1} for a bipartite state that, on average,  $H_{A|B}$ will increase under any LOCC.  In particular, we wish to prove that for a given state and any two LOCC operators $L_A$ and $L_B$,
\begin{equation}
|\psi'  \rangle =  M_{AB} |\psi \rangle = L_A \otimes L_B |\psi \rangle,
\end{equation} 
that both the entropy and distance increase,
\begin{equation}
\begin{split}
&H_{A|B}(\psi') \geq H_{A|B}(\psi) \\
&D_{AB}(\psi') \geq  D_{AB}(\psi);
\end{split}
\end{equation}
respectively.
Consider the local operations acting on the density matrix
\begin{equation}
M \rho M^\dag = M |\psi \rangle \langle \psi | M^\dag,
\end{equation}  
then necessarily 
\begin{equation}
\rho'= \sum_{i,j,k,l} Tr \left(M_{\!{}_{A_i\!B_j}}\,  \rho\,  M^\dag_{\!{}_{A_k\!B_l}}\right) \left[M_{\!{}_{A_i\!B_j}}\,  \rho\,  M^\dag_{\!{}_{A_k\!B_l}}\right]
\end{equation}
Using the convexity of Shannon entropy we have
\begin{equation}
H_{A|B}(\rho)\geq \sum_{i,j,k,l}  Tr \left(M_{\!{}_{A_i\!B_j}}\,  \rho\,  M^\dag_{\!{}_{A_k\!B_l}}\right) H_{AB} \left(M_{\!{}_{A_i\!B_j}}\,  \rho\,  M^\dag_{\!{}_{A_k\!B_l}}\right)
\end{equation}
Then $\mathcal{R}(\psi')\leq \mathcal{R}(\psi)$.
Although all the proofs in this section are written for two qu$b$it systems they are easily generalized to a larger number of qu$b$its.   Since in in higher dimensions the reactivity for $n$ observers, $\{C_i\}_{i,1,2,\ldots,n}$ will still satisfy
\begin{equation}
{\mathcal R} \propto \frac{1}{\overline H_{C_1 C_2 \ldots  C_n}}.
\end{equation}

\subsection {Reactivity is Monotonic, its Maximum Bound Requires Quantum Correlations.}
\label{sec:property3}
It is well established  that quantum correlation is a resource for quantum processing.\cite{Preskill:2012}  The correlation in entangled quantum states is stronger than any classical correlation.  Unlike classical correlation, quantum correlation is non-vanishing in more than one basis. 
In particular, a bipartite state with large quantum correlation may yield the same information distance ${\mathcal D}_{AB}$ as classically correlated quantum state for a given measurement operator $M_{AB}$. This is not acceptable for a measure of quantum correlation.  However,  our definition of reactivity in Eqs.~\ref{eq:R1}\&\ref{Eq:C} averages over all measurements and the information distance will then be larger for the classically--correlated quantum state.  Therefore, the reactivity of the quantum correlated system will be larger since its inversely proportional to the information distance.
 For example,  even though  the two states,
 \begin{equation}
 |\psi \rangle = |HH \rangle\ \ \hbox{(no quantum correlation)}
 \end{equation} 
 and
 \begin{equation}
 |\psi' \rangle = \frac{|HH \rangle + |VV \rangle}{\sqrt{2}}\ \ \hbox{(maximal quantum correlation)}
 \end{equation}
 have the same distance $D_{AB}=D'_{AB}$ for Alice and Bob given the horizontal--horizontal ($HH$) measurement basis measurement; nevertheless,  when we take an average over all configurations of Alice and Bob we showed that  
 \begin{equation}
 \overline{D_{AB}}(\psi) \geq  \overline{D_{AB}}(\psi')
 \end{equation}
then maximum bound created by maximally entangled state can't be created by any classical correlation.

\section{Summary of the Properties of Reactivity}
\label{fini}
We showed in this paper that the geometrically-defined reactivity over the space of measurements satisfies the major properties required of a measure of quantum correlation. The reactivity measure is scalable in the sense that it can be generalized to higher number of qu$b$its.  As a measure of correlation it has the advantage of being interpretation free unlike quantum discord for multipartite states.  Its expression is a relatively straightforward analytic function probabilities, and it does not require any global minimization procedure or matrix inversion.  In other words, it appears to us to be relatively  easy to calculate in comparison to other measures of correlation.   Nevertheless, its computational complexity is driven by the need to compute joint entropies over the observers measurement outcomes.  For a qudit state this would require a $d$--fold summation that scales exponentially.   However, we may be able to extract an accurate measure of the correlation for a multipartite system with high fidelity by using only a fraction of the possible measurements.  An analysis of the fidelity of this and other entanglement measures under partial--measures, and its impact on reducing the computational complexity is beyond the scope of this paper.   Nevertheless,  we are animated by Quantum Sanov's Theorem that shows that the fidelity of distinguishing two quantum density matrices pure $\rho_2$ from $\rho_1$ improves exponentially with the number of measurements, $N$,\cite{Vedral:1997}
\begin{equation}\label{eq:Sanov}
\left(
\begin{array}{c}
Fidelity\ of\ \rho_1 \rightarrow \rho_2\\
with\ N\ measurements
\end{array}
\right)
= 1-e^{-N S\left(\rho_1||\rho_2\right)}.
\end{equation} 
Here, $S(\rho_1||\rho_2) = Tr(\rho_1 \log \rho_1 - \rho_1 \log \rho_2)$ is a relative entropy.  We hope that this theorem's ``information thermodynamic" structure can be extended to a larger classes of quantum states. 

\section*{Acknowledgment}
In  here we thank David Meyer and Alexander Meill  from UCSD for probing questions on our previous paper that motivated us to expound on the four properties of the mean reactivity.  PMA and WAM would like thank support from the Air Force Office of Scientific Research (AFOSR).  WAM  research was supported under AFOSR/AOARD grant \#FA2386-17-1-4070.  One of us (WAM) would also like to acknowledge the support from the Griffiss Institute and the Air Force Research Laboratory at Rome Labs under the Visiting Faculty Research Program.  Any opinions, findings, conclusions or recommendations expressed in this mate- rial are those of the author(s) and do not necessarily reflect the views of AFRL.

\bibliography{qig4-2019} 
\bibliographystyle{plain} 

\end{document}